\documentclass[aps,pra,twocolumn,showpacs,superscriptaddress]{revtex4}
\usepackage[pdfview=FitV,pdfstartview=FitV]{hyperref}
\usepackage{amsmath,amssymb,amsfonts,empheq,bm}
\usepackage{graphicx}
\usepackage{epsfig}
\usepackage{color}
\newcommand{\mbf}[1]{{\textbf #1}}

\begin{document}
\title{Time-of-Flight Roton Spectroscopy in Dipolar Bose-Einstein Condensates}

\author{M. Jona-Lasinio}
\email{mattia.jonalasinio@roma1.infn.it}
\affiliation{Institut f\"ur Theoretische Physik, Leibniz Universit\"at, 30167
Hannover, Germany}
\affiliation{Dipartimento di Fisica, Universit\`a di Roma ``La Sapienza'', 00185
Roma, Italy}

\author{K. {\L}akomy}
\affiliation{Institut f\"ur Theoretische Physik, Leibniz Universit\"at, 30167
Hannover, Germany}

\author{L. Santos}
\affiliation{Institut f\"ur Theoretische Physik, Leibniz Universit\"at, 30167
Hannover, Germany}

\date{\today}

\begin{abstract}
Dipolar Bose-Einstein condensates may present a rotonlike dispersion minimum, 
which has yet to be observed in experiments. We discuss a simple method to 
reveal roton excitations, based on the response of quasi-two-dimensional dipolar
condensates against a weak lattice potential. By employing numerical 
simulations for realistic scenarios, we analyze the response of the system as a
function of both the lattice spacing and the $s$-wave scattering length, 
showing that the roton minimum may be readily revealed in current experiments 
by the resonant population of Bragg peaks in time-of-flight measurements.
\end{abstract}

\maketitle


{\it Introduction.} Dipole-dipole interactions in ultracold gases result in a
rich physics extensively investigated in recent 
years~\cite{Baranov2008,Lahaye2009,Baranov2012}. Quantum degeneracy has been 
achieved already in gases of highly magnetic atoms such as 
chromium~\cite{Griesmaier2005,Beaufils2008}, dysprosium~\cite{Lu2011,Lu2012}, 
and erbium~\cite{Aikawa2012}. In addition, ultracold samples of heteronuclear
molecules in their rovibrational ground state allow for the creation of a
molecular degenerate polar gas~\cite{Ni2008,DeMiranda2011,Chotia2012}.


The interesting features of ultracold polar gases originate from the anisotropic
and long-range character of the dipole-dipole
interaction~\cite{Baranov2008,Lahaye2009,Baranov2012} that results in a
geometry-dependent stability~\cite{Muller2011} and in a peculiar dispersion of
the elementary excitations. Gases with short-range interactions present a linear
(phonon) dispersion at low momenta, monotonically increasing to a
quadratic~(free particle) dispersion at large momenta~\cite{Stringari-Book}. In
contrast, under proper conditions, dipolar Bose-Einstein condensates present a
dispersion minimum at intermediate momenta~\cite{Santos2003,Ronen2007}, a
feature reminiscent of the roton minimum of superfluid
helium~\cite{Landau1947,Feynman1954}. Moreover, when the rotonlike minimum
reaches zero energy, the condensate becomes dynamically unstable (roton
instability)~\cite{Santos2003,Ronen2007,Wilson2009,JonaLasinio2013} against
finite momentum excitations, and it collapses in a fundamentally different way
compared to the usual phonon collapse~\cite{Lahaye2008}. 


Although rotons constitute a key feature of dipolar gases, the direct 
observation and characterization of rotonlike excitations in these gases is 
still elusive, being nowadays a major goal pursued by several experimental 
groups~\cite{Wu2012,Takekoshi2012}. The detection of rotons is complicated by 
several issues. On one hand, the rotonized spectrum results from the peculiar 
momentum dependence of the dipole-dipole interactions in quasi-two-dimensional
(or quasi-one-dimensional) traps, with a strong transverse harmonic confinement.
The characteristic roton wavelength is approximately $l_z$, where $l_z$ is the
transverse oscillator length. As a result, the observation of the roton requires
pancake-like traps with large aspect ratios. On the other hand, harmonically
confined pancake condensates with large aspect ratios are characterized by an
inherent local nature of the roton excitation spectrum, which leads to an
effective roton confinement at the center of the
trap~\cite{JonaLasinio2013,Bisset2013} and therefore to a reduced roton signal.

Various methods have been proposed so far to detect the rotonized spectrum. For
example the structure factor of a trapped dipolar gas was studied recently in
Ref.~\cite{Blakie2012}. However, the local nature of the roton excitations may
hinder the clear characterization of the position of the dispersion minimum. In
another recent interesting paper~\cite{Bisset2013}, it has been shown that
atom-number fluctuations may be employed to reveal not only the presence of the
roton minimum but also its local confinement. However, this technique requires
{\it in situ} imaging~\cite{footnote-Axel}, which still represents a serious
technical challenge in current experiments. As a third way, rotons may be
detected by stability spectroscopy~\cite{Corson2013a,Corson2013b} based on the
analysis of the condensate stability in the presence of a weak lattice. This
technique, which has been analyzed in the absence of an overall harmonic
confinement, may still be handicapped by the local nature of the roton spectrum
in trapped condensates since the very presence of the confinement leads to a
local, rather than global, stability of the condensate. As a consequence the
onset of destabilization will also be local, thus impeding the precise
determination of the roton wavelength and the roton instability regime.

In this Brief Report we propose a simple technique to reveal the roton, which 
similar to Refs.~\cite{Corson2013a,Corson2013b} is based on the response of  the
condensate against a weak lattice potential. However, we do not focus on  the
stability of the condensate but rather on the effects of the rotonlike  minimum
on time-of-flight pictures of the stable condensate. The presence of a deep
roton minimum at the center of the trap results in an enhanced local  response
against a weak lattice potential. The magnitude of the Bragg peaks in
time-of-flight images presents a resonant dependence as a function of the 
lattice spacing, peaking in the vicinity of the roton wavelength. As such, a
clear signature of the roton minimum can be revealed without {\it in situ}
imaging but simply with time-of-flight images. We analyze this resonant
dependence and the variation of the Bragg peak by means of numerical
simulations, as a function of both the $s$-wave scattering length and the
lattice spacing. We carefully consider the effects of confinement by simulating
realistic conditions in on-going experiments, showing that the Bragg peak may be
readily resolved, hence opening a promising route for the observation of the
roton minimum.

\begin{figure}[t]
\centering
\includegraphics[clip=true,width=1.0\columnwidth]{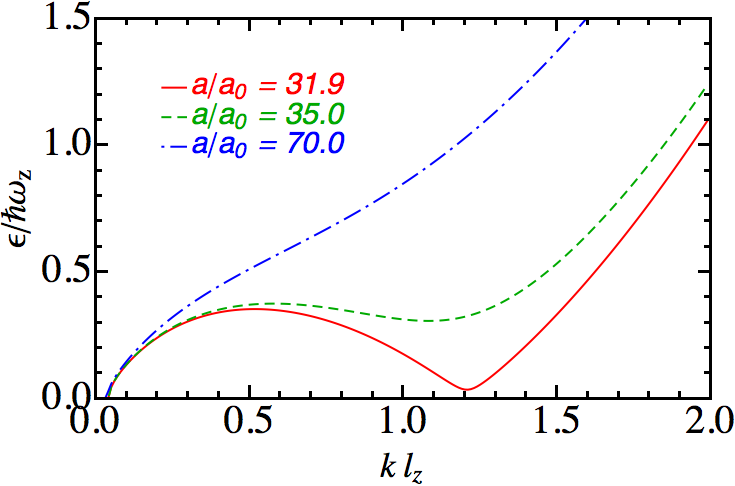}
\vspace*{-0.2cm}
\caption{
(Color online) Dispersion $\epsilon(k)$ for a dipolar Bose-Einstein condensate
of $N=10^5$ erbium atoms with trapping parameters $\omega=2\pi \times 50$ Hz,
$\omega_z=2\pi \times 700$ Hz without lattice potential ($V_0=0$; see text). We
identify the deep roton region at $a=31.9\, a_0$ (red solid line) close to the
stability threshold $a_{\rm crit} = 31.6\, a_0$, the shallow roton region at
$a=35.0\, a_0$ (green dashed line), and the monotonic region at $a=70.0\, a_0$
(blue dash-dotted line).
}
\label{fig:1}
\end{figure}

{\it Model.} We consider a dipolar condensate of $N$ bosons of mass $m$ and
(electric or magnetic) dipole moment $d$ oriented along $z$. The condensate is
confined by a three-dimensional (3D) harmonic trap of frequencies $\omega$ in
the $xy$ plane and $\omega_z=\lambda \omega$ along $z$, with $\lambda\gg 1$.
Within a mean field approach, the condensate wave function $\psi(\mbf{r},t)$
obeys the nonlocal Gross-Pitaevskii equation,
\begin{multline}
\label{eq:GP}
i\hbar\frac{\partial}{\partial t}\psi(\mbf{r},t)=
\left[ 
-\frac{\hbar^2\nabla^2}{2m}+ \frac{m\omega^2}{2}\left(\rho^2+\lambda^2 z^2
\right) + V_{\rm lat}(\mbf{r}) + \right. \\ 
 \left. + g|\psi(\mbf{r},t)|^2 + 
 \int d^3 r' V_{dd}(\mbf{r}-\mbf{r'}) |\psi(\mbf{r'},t)|^2
\right]
 \psi(\mbf{r},t), 
\end{multline}
where $g=4\pi\hbar^2 aN/m$ characterizes the short-range interaction, $a$ is the
$s$-wave scattering length, $V_{dd}(\mbf{r})=N\,d^2\,(1-3\cos^2\theta)/r^3$ is
the dipole-dipole interaction potential, $\theta$ is the angle between $\mbf{r}$
and the $z$ axis, $\rho$ is the radial coordinate in the $xy$ plane and $\int
d^3 r |\psi(\mbf{r},t)|^2=1$.

We consider a weak perturbing one-dimensional (1D) lattice formed by two
intersecting lasers of wavelength $\lambda_0$, propagating on the $yz$ plane
with wave vectors ${\bf k}_{1,2}=\frac{2\pi}{\lambda_0}(\pm \cos\alpha {\bf
e}_y+\sin\alpha {\bf e}_z)$, where $\alpha$ is the intersection angle. The
resulting lattice potential acquires the form $V_{\rm
lat}(\mbf{r})=V_0\sin^2\left(k_L y / 2\right)$, with a lattice spacing
$2\pi/k_L=\lambda_0/2\cos\alpha$. Note that by changing the angle $\alpha$ we
can explore all lattice spacings larger than $\lambda_0/2$, an important feature
for the experimental implementation of our method. 

As mentioned above, a pancake dipolar condensate may become unstable against 
roton instability at intermediate momenta $k \sim 1/l_z$~\cite{Santos2003}, 
with $l_z$ being the transverse oscillator wavelength. This occurs for an
$s$-wave  scattering $a<a_{\rm crit}$, where the value of the critical
scattering length $a_{\rm crit}$ depends on the dipole strength and on the
confinement. In the presence of the weak additional lattice, $a_{\rm crit}$ also
depends on the lattice properties. For a given $V_0$ we determine $a_{\rm crit}$
first. Then, for each particular lattice spacing $2\pi/k_L$ and each scattering
length $a > a_{\rm crit}$, we calculate the ground state $\psi_0(\mbf{r})$ of
the system, using the imaginary time evolution of Eq.~\eqref{eq:GP}.

We are particularly interested in the properties of the density modulations 
induced by the weak lattice. In general, a lattice with amplitude $V_0$ smaller
than the chemical potential leads to weak or vanishing density modulations. 
However, this is not the case in the presence of a deep roton minimum in the 
spectrum. To analyze the response of the system against the perturbing lattice,
we obtain the integrated two-dimensional (2D) density $n^{2D}(x,y)=\int dz\,
|\psi_0(\mbf{r})|^2$ and calculate the visibility $c$ of the associated density
pattern, defined as
\begin{equation}
c=\frac{n^{2D}_{\rm MAX}-n^{2D}_{\rm min}} {n^{2D}_{\rm MAX}+n^{2D}_{\rm min}},
\label{eq:c}
\end{equation}
where $n^{2D}_{\rm MAX}$ is the central maximum of the density modulation 
while $n^{2D}_{\rm min}$ is the adjacent minimum. For the considered 1D 
lattice we have $n^{2D}_{MAX} \equiv n^{2D}(0,0)$ and $n^{2D}_{min} \equiv 
n^{2D}(0,\pi/k_L)$.

{\it Numerical results.} Our calculations are performed considering a 
condensate of erbium, but similar results hold for other magnetic atoms, 
although the particular parameters will, of course, be different. Based on 
ongoing experiments, we consider the particular case of $N=10^5$ $^{168}$Er 
atoms in a cylindrical trap with frequencies $\omega_z = 2\pi \times 700$ Hz and
$\omega = 2\pi \times 50$ Hz ($\lambda = 14$). We consider a lattice strength
$V_0 = 0.34 \, \hbar \omega_z$. The chemical potential $\mu$ weakly  depends on
$k_L$, but we checked that $V_0 \lesssim 0.1\mu$ for all the lattice momenta we
consider. For these parameters we determine $a_{\rm crit} = 38.0\, a_0$, where
$a_0$ is the Bohr radius. In Fig.~\ref{fig:1} we plot the Bogoliubov spectrum of
the condensate, obtained as the solution of the Bogoliubov-de Gennes equations
for the case $V_0=0$, as detailed in Refs.~\cite{Santos2003,JonaLasinio2013}. In
this particular case we find $a_{\rm crit}=31.6\, a_0$ since the perturbing
lattice $V_{\rm lat}$ obviously has a destabilizing effect on the condensate. By
properly tuning the scattering length using Feshbach resonances it is possible
to explore the different roton regimes illustrated in Fig.~\ref{fig:1}: deep
roton, $a \sim a_{\rm crit}$; shallow roton $a > a_{\rm crit}$; and roton free,
$a \gg a_{\rm crit}$.

\begin{figure}[t]
\centering
\includegraphics[clip=true,width=1.0\columnwidth]{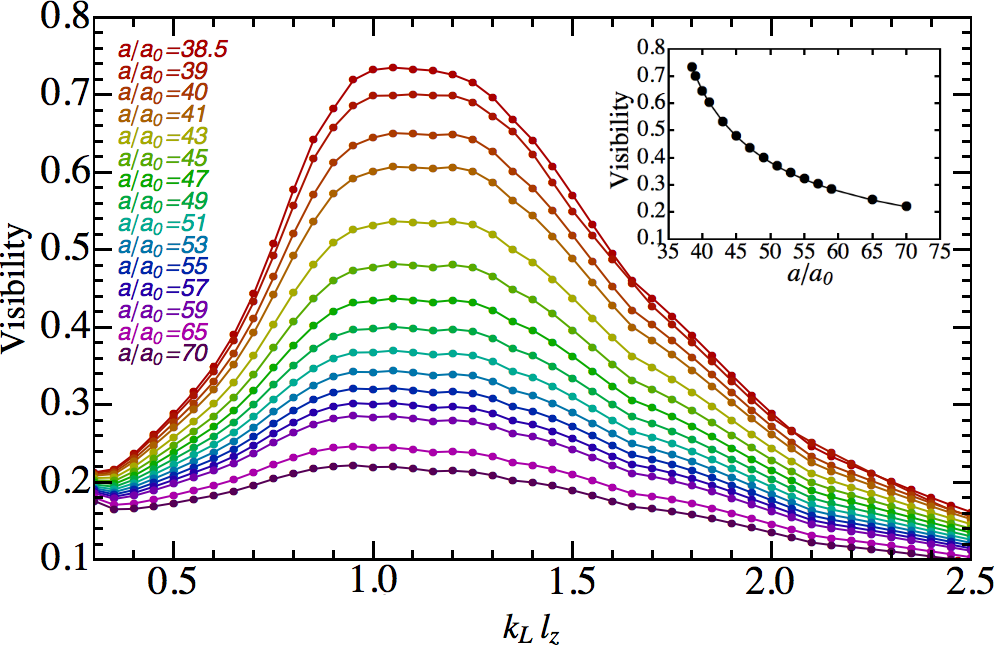}
\vspace*{-0.2cm}
\caption{
(Color online) Visibility of the imprinted density pattern for $N=10^5$ erbium
atoms as a function of $k_L$. Parameters are $\omega=2\pi \times 50$ Hz,
$\omega_z=2\pi \times 700$ Hz ($\lambda=14$) and lattice depth
$V_0=0.34\,\hbar\omega_z$ ($V_0 \lesssim 0.1 \mu$; see text). The inset shows
the visibility as a function of the scattering length $a$ for $k_L l_z=1$.
}
\label{fig:2}
\end{figure}

\begin{figure}[t]
\centering
\includegraphics[clip=true,width=1.0\columnwidth]{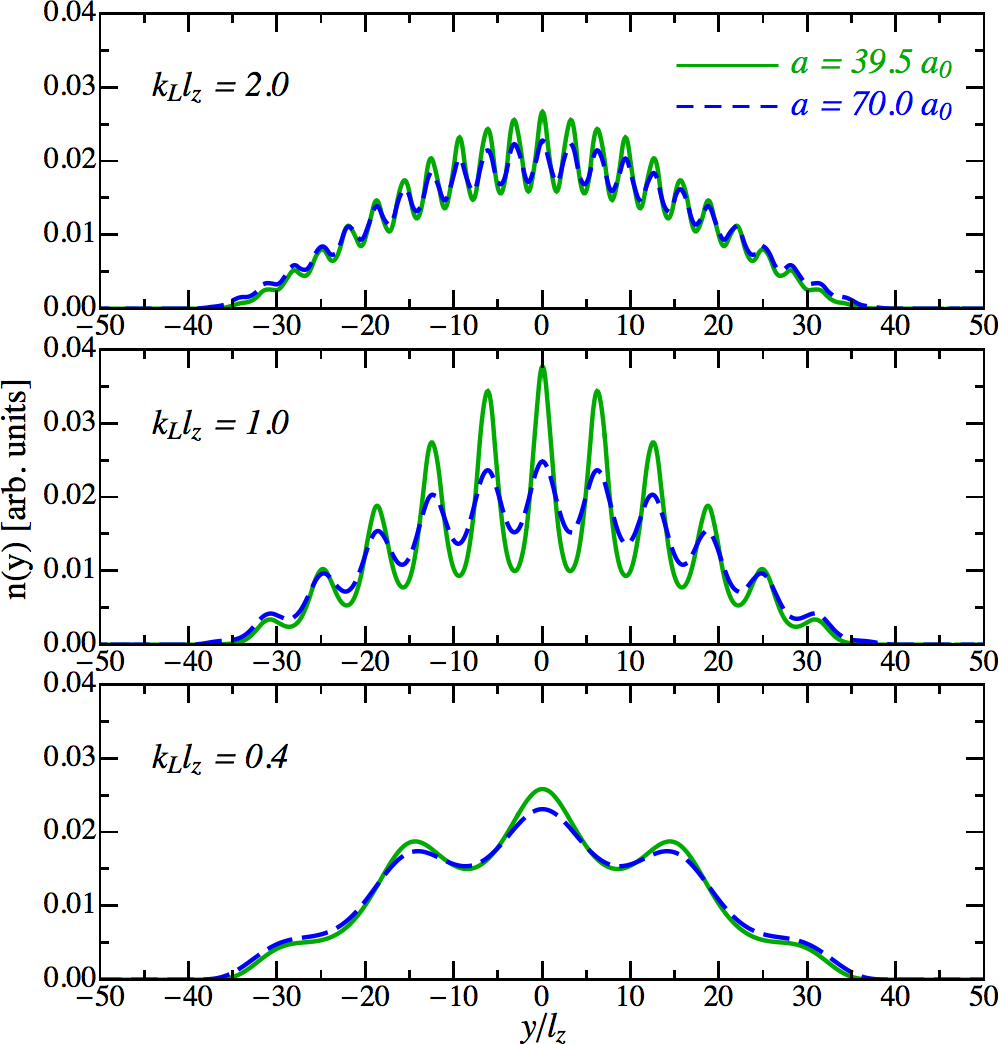}
\vspace*{-0.2cm}
\caption{
(Color online) Integrated spatial distribution $n(y)$ (see text) deep in the
roton regime at $a=39.5\,a_0$ (green solid line) and far from the roton regime
at $a=70.0\,a_0$ (blue dashed line), for $k_L l_z$ = 0.4 (bottom panel), 1.0
(middle panel), 2.0 (top panel).
}
\label{fig:3}
\end{figure}

We consider the scattering length interval $38.5 \le a/a_0 \le 70.0$ and the
momentum interval $0.3 \le k_L l_z \le 2.5$. By changing the intersection angle
$\alpha$, we can explore all values $k_L l_z <4\pi l_z/\lambda_0$. Note that for
the parameters discussed above $l_z\simeq 300$ nm, and hence for
$\lambda_0\simeq 500$nm, one would be able to scan all $k_L l_z <7.5$. Note,
however, that for $k_L l_z < 0.3$ the lattice wavelength is comparable to the
radial size of the condensate, and the concept of visibility loses its meaning.
The visibility $c$ as a function of the lattice momentum $k_L$ is plotted in
Fig.~\ref{fig:2} for several values of the scattering length $a$. For $a \sim
70\, a_0$, far away from the deep roton regime, the visibility is approximately
constant around $c\sim 0.2$, being basically independent of $k_L$. In contrast,
for $a \sim 38.5\, a_0$, close to roton instability, the contrast of the density
pattern presents a marked resonance-like dependence on $k_L$, increasing very
significantly from $c\lesssim 0.2$ to $c \sim 0.75$, around $k_L l_z\sim 1.1$.
As shown in Fig.~\ref{fig:1} this value coincides with the expected roton
wavelength. Hence, by measuring the contrast of the density pattern, we have
access to a conclusive signature of the roton, as well as to a precise
determination of the roton wavelength. This technique would, however, still
require {\it in situ} measurements~\cite{footnote-Axel}. In Fig.~\ref{fig:3} we
plot the integrated density distribution $n(y) = \int dz\, dx\,
|\psi_0(\mbf{r})|^2$ along the lattice direction in real space to illustrate how
the contrast changes as a function of the lattice momentum and of the scattering
length.

To avoid the challenging requirement of {\it in situ} imaging we now analyze the
effects of the presence of a rotonlike minimum in time-of-flight pictures. The
density modulation discussed above results in a momentum distribution
characterized by a central peak at $k_y \sim 0$, corresponding to the
low-momentum components, and two side Bragg peaks at $k_y \sim \pm k_L$
associated with the lattice recoil momentum. This momentum distribution may be
imaged in real space after a time-of-flight expansion. A caveat is, however,
necessary at this point. As has been shown recently, the time-of-flight
expansion of dipolar gases in deep optical lattices may trigger the collapse of
the condensate even if the condensate was stable in the trap~\cite{Billy2012}.
However, we always consider here a weak lattice strength compared to the 3D
chemical potential $\mu$, so that all the interference peaks are stable against
collapse and separate during the time-of-flight expansion.

\begin{figure}[t]
\centering
\includegraphics[clip=true,width=1.0\columnwidth]{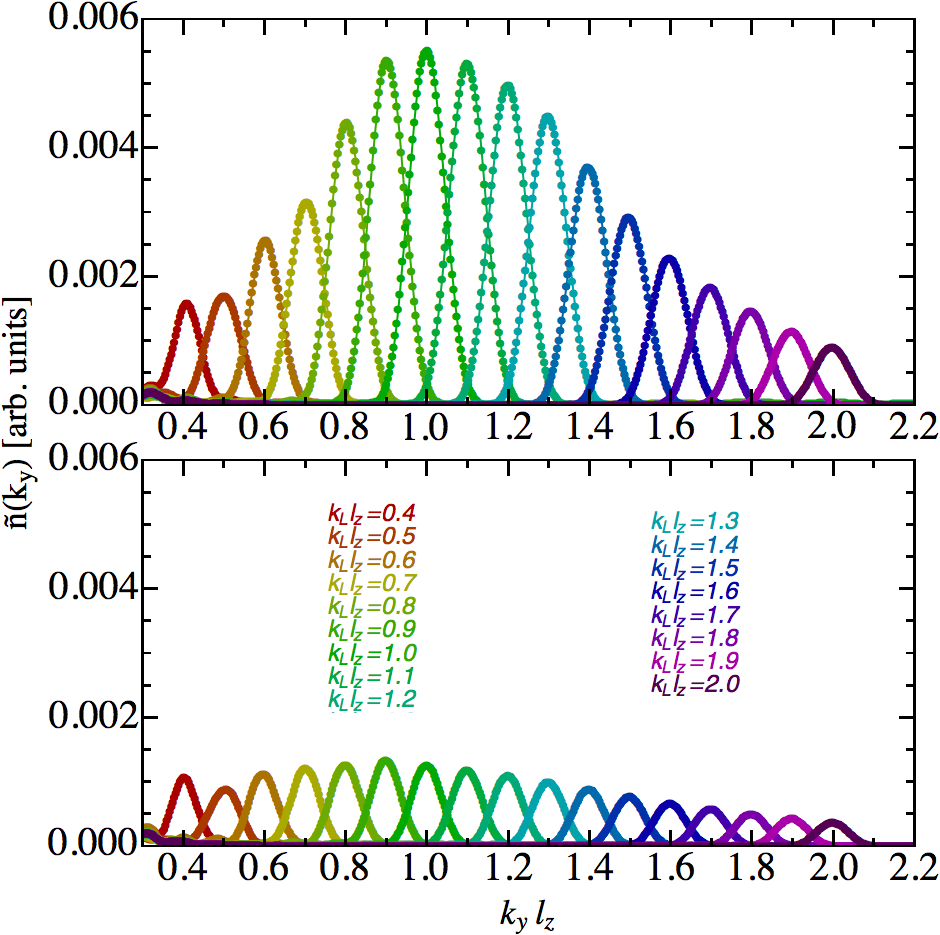}
\vspace*{-0.2cm}
\caption{
(Color online) Integrated momentum distribution $\tilde n(k_y)$~(see text) deep
in the roton regime at $a=39.5\, a_0$ (top panel) and far from the roton regime
at $a=70.0\, a_0$ (bottom panel) for $0.4 \le k_L l_z \le 2.0$, with other
parameters being the same as those in Fig.~\ref{fig:2}. The envelope of the
Bragg peaks reveals the enhanced sensitivity at $k_L l_z \sim 1.1$.}
\label{fig:4}
\end{figure}

In order to address the results of a time-of-flight measurement, we calculate
the integrated momentum distribution of the condensate along the lattice
direction, $\tilde n(k_y) =\int \frac{dk_z}{2\pi}\,\frac{dk_x}{2\pi} |\tilde
\psi (\mbf{k})|^2$, for different values of the lattice momentum $k_L$; see
Fig.~\ref{fig:4}. In the parameter region where the dispersion of the condensate
shows a roton minimum we observe an increase of the Bragg peak at $k_L$ whenever
the lattice momentum matches that of the condensate roton
mode~(Fig.~\ref{fig:4}, top panel, $a=39.5\, a_0$). In contrast, if the roton
minimum is very shallow or absent ~(Fig.~\ref{fig:4}, bottom panel, $a=70.0\,
a_0$), the magnitude of the Bragg peak is almost insensitive to the lattice
momentum. The envelope of all the Bragg peaks for different $k_L$
(Fig.~\ref{fig:4}) nicely illustrates the enhanced sensitivity.

\begin{figure}[t]
\centering
\includegraphics[clip=true,width=1.0\columnwidth]{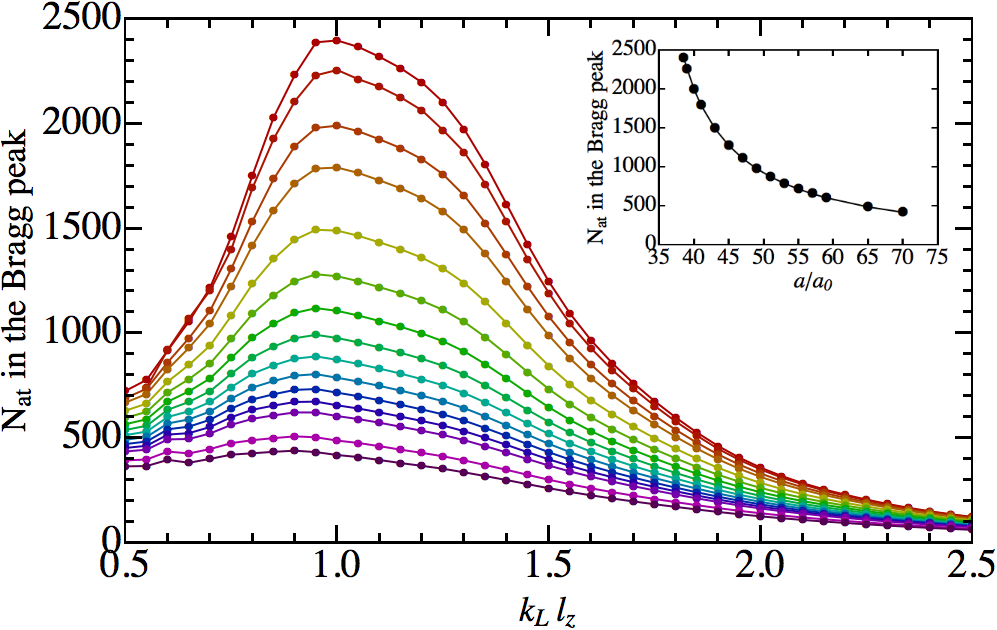}
\vspace*{-0.5cm}
\caption{
(Color online) Number of atoms in a single Bragg peak for the parameters and
color code in Fig.~\ref{fig:2}. The inset shows the number of atoms in a single
Bragg peak as a function of the scattering length $a$ for $k_L l_z=1$.
}
\label{fig:5}
\end{figure}

In Fig.~\ref{fig:5} we plot the absolute number of atoms in a single Bragg peak,
assuming $N=10^5$ atoms in the condensate. The number of atoms in a single Bragg
peak reaches its maximum around $2400$. Note that in spite of the large maximal
contrast, $c\simeq 0.75$, the corresponding population of each Bragg peak is at
most $2.4\%$ of the total number of atoms. This may be understood by noting that
Bragg peaks are in general weakly sensitive to amplitude modulations. A
homogeneous system with a modulation $\psi(x)=\psi_0 [1+A\cos(kx)]$ presents a
visibility $c=2A/(1+A^2)$, whereas the relative weight of a single Bragg peak is
$\eta=A^2/(2+A^2)$. A contrast of $c=0.75$ hence corresponds to $\eta=0.092$.
Second, the number of particles in the Bragg peak is further reduced by the
local nature of the roton spectrum since the density modulations are only
markedly contrasted at the center of the condensate. This explains the
additional reduction of the maximal number of atoms at each Bragg peak,
resulting in a final $2.4\%$ of the overall number, well below the population of
the central peak. Note, however, that the Bragg peaks clearly separate from the
dominant zero-momentum component during the time-of-flight expansion, and hence
they may still be imaged without the perturbing effect of the dominant
low-momentum peak. Note as well that although the relative percent is low, the
absolute number of atoms in the Bragg peaks is certainly sizable. Numbers over
$1000$ atoms are resolvable in current experiments. In this sense, we point out
that the Bragg peak is narrowly concentrated around $(k_x=0, k_y=k_L)$, and
hence results in a narrow density peak in the time-of-flight expansion, which
will consequently result in an enhanced signal at the detectors. Therefore by
measuring the number of atoms in the Bragg peaks for different values of the
scattering length it should be possible to conclusively reveal roton
excitations.

{\it Conclusion.} We have proposed a simple way to reveal roton excitations in
quasi-2D trapped dipolar condensates based on the response of the  condensate in
the presence of a weak lattice. We have shown that even a weak  lattice may
induce a large density modulation at the center of the condensate. This
modulation may be revealed by a sizable Bragg peak in time-of-flight  pictures,
therefore avoiding {\it in situ} imaging. We have analyzed the properties of
both the contrast of the density modulation and the magnitude of the associated
Bragg peak, showing that they present a resonant character as a function of the
lattice spacing, peaking around the roton wavelength. The resonant behavior
becomes more and more marked as the condensate approaches the roton instability.
Finally, we have analyzed a realistic scenario based on current erbium
experiments, showing that this technique allows for a conclusive observation of
the roton excitations under current experimental conditions.

\acknowledgments
We acknowledge funding from the German-Israeli Foundation and the DFG (Grant
No. SA1031/6 and Excellenzcluster QUEST).


\begin{thebibliography}{99}

\bibitem{Baranov2008} M. A. Baranov, Phys. Rep. {\bf 464}, 71 (2008).

\bibitem{Lahaye2009} T. Lahaye, C. Menotti, L. Santos, M. Lewenstein, and T. Pfau, Rep. Prog. Phys. {\bf 72}, 126401 (2009). 

\bibitem{Baranov2012} M. A. Baranov, M. Dalmonte, G. Pupillo, and P. Zoller, Chem. Rev. {\bf 112}, 5012 (2012).

\bibitem{Griesmaier2005} A. Griesmaier, J. Werner, S. Hensler, J. Stuhler, and T. Pfau, Phys. Rev. Lett. {\bf 94}, 160401 (2005).

\bibitem{Beaufils2008} Q. Beaufils, R. Chicireanu, T. Zanon, B. Laburthe-Tolra, E. Mar\'echal, L. Vernac, J.-C. Keller, and O. Gorceix, Phys. Rev. A {\bf 77}, 061601 (2008).

\bibitem{Lu2011} M. Lu, N. Q. Burdick, S. H. Youn, and B. L. Lev, Phys. Rev. Lett. {\bf 107}, 190401 (2011). 

\bibitem{Lu2012} M. Lu, N. Q. Burdick, and B. L. Lev, Phys. Rev. Lett {\bf 108}, 215301 (2012).

\bibitem{Aikawa2012} K. Aikawa, A. Frisch, M. Mark, S. Baier, A. Rietzler, R. Grimm, and F. Ferlaino, Phys. Rev. Lett. {\bf 108}, 210401 (2012).

\bibitem{Ni2008} K.-K. Ni, S. Ospelkaus, M. H. G. de Miranda, A. Pe'er, B. Neyenhuis, J. J. Zirbel, S. Kotochigova, P. S. Julienne, D. S. Jin, and J. Ye, Science {\bf 322}, 231 (2008).

\bibitem{DeMiranda2011} M. H. G. de Miranda, A. Chotia, B. Neyenhuis, D. Wang, G. Qu\'em\'ener, S. Ospelkaus, J. L. Bohn, J. Ye, and D. S. Jin, Nat. Phys. {\bf 7}, 502 (2011).

\bibitem{Chotia2012} A. Chotia, B. Neyenhuis, S. A. Moses, B. Yan, J. P. Covey, M. Foss-Feig, A. M. Rey, D. S. Jin, and J. Ye, Phys. Rev. Lett. {\bf 108}, 080405 (2012).

\bibitem{Muller2011} T. Koch, T. Lahaye, J. Metz, B. Fr\"ohlich, A. Griesmaier, and T. Pfau, Nature Phys. {\bf 4}, 218 (2008);
S. M\"uller, J. Billy, E. A. L. Henn, H. Kadau, A. Griesmaier, M. Jona-Lasinio, L. Santos, and T. Pfau, Phys. Rev. A {\bf 84}, 053601 (2011).

\bibitem{Stringari-Book} L. Pitaevskii and S. Stringari, {\it Bose-Einstein Condensation}, (Oxford University Press, New York, 1993).

\bibitem{Santos2003} L. Santos, G. V. Shlyapnikov, and M. Lewenstein, Phys. Rev. Lett. {\bf 90}, 250403 (2003).

\bibitem{Ronen2007} S. Ronen, D. C. E. Bortolotti, and J. L. Bohn, Phys. Rev. Lett. {\bf 98}, 030406 (2007).

\bibitem{Landau1947} L. D. Landau, J. Phys. USSR {\bf 11}, 91 (1947); Phys. Rev. {\bf 75}, 884 (1949).

\bibitem{Feynman1954} R. P. Feynman, Phys. Rev. {\bf 94}, 262 (1954).

\bibitem{Wilson2009} R. M. Wilson, S. Ronen, and J. L. Bohn, Phys. Rev. A {\bf 80}, 023614 (2009).

\bibitem{JonaLasinio2013} M. Jona-Lasinio, K. {\L}akomy and L. Santos, Phys. Rev. A {\bf 88}, 013619 (2013).

\bibitem{Lahaye2008} T. Lahaye, J. Metz, B. Fr\"olich, T. Koch, M. Meister, A. Griesmaier, T. Pfau, H. Saito, Y. Kawaguchi, and M. Ueda, Phys. Rev. Lett. {\bf 101}, 080401 (2008).

\bibitem{Wu2012} C.-H. Wu, J. W. Park, P. Ahmadi, S. Will, and M. W. Zwierlein, Phys. Rev. Lett. {\bf 109}, 085301 (2012). 

\bibitem{Takekoshi2012} T. Takekoshi {\it et al.}, Phys. Rev. A {\bf 85}, 032506 (2012).

\bibitem{Bisset2013} R. N. Bisset and P. B. Blakie, Phys. Rev. Lett. {\bf 110}, 265302 (2013).

\bibitem{Blakie2012} P. B. Blakie, D. Baillie, and R. N. Bisset, Phys. Rev. A {\bf 86}, 021604 (2012).

\bibitem{footnote-Axel} {\it In situ} imaging may be possible in on-going dysprosium experiments. A. Griesmaier (private communication). 

\bibitem{Corson2013a} J. P. Corson, R. Wilson and J. L. Bohn, Phys. Rev. A {\bf 87}, 051605 (2013).

\bibitem{Corson2013b} J. P. Corson, R. Wilson and J. L. Bohn, Phys. Rev. A {\bf 88}, 013614 (2013).

\bibitem{Billy2012} J. Billy, E. A. L. Henn, S. M\"uller, T. Maier, H. Kadau, A. Griesmaier, M. Jona-Lasinio, L. Santos, and T. Pfau, Phys. Rev. A {\bf 86}, 051603 (2012).

\end{thebibliography}
\end{document}